\documentclass[12pt]{article}
\usepackage{a4wide}
\usepackage{amssymb}
\usepackage{graphicx}
\begin{document}
{\renewcommand{\thefootnote}{\fnsymbol{footnote}}
\begin{center}
{\LARGE Abelianized structures in spherically symmetric hypersurface
  deformations}\\
\vspace{1.5em}
Martin Bojowald\footnote{e-mail address: {\tt bojowald@psu.edu}}
\\
\vspace{0.5em}
Institute for Gravitation and the Cosmos,\\
The Pennsylvania State
University,\\
104 Davey Lab, University Park, PA 16802, USA\\
\vspace{1.5em}
\end{center}
}

\setcounter{footnote}{0}

\begin{abstract}
  In canonical gravity, general covariance is implemented by
  hypersurface-deformation symmetries on phase space. The different versions
  of hypersurface deformations required for full covariance have complicated
  interplays with one another, governed by non-Abelian brackets with structure
  functions. For spherically symmetric space-times, it is possible to identify
  a certain Abelian substructure within general hypersurface deformations,
  which suggests a simplified realization as a Lie algebra. The generators of
  this substructure can be quantized more easily than full hypersurface
  deformations, but the symmetries they generate do not directly correspond to
  hypersurface deformations. The availability of consistent quantizations
  therefore does not guarantee general covariance or a meaningful quantum
  notion thereof. In addition to placing the Abelian substructure within the
  full context of spherically symmetric hypersurface deformation, this paper
  points out several subtleties relevant for attempted applications in
  quantized space-time structures. In particular, it follows that recent
  constructions by Gambini, Olmedo and Pullin in an Abelianized setting fail
  to address the covariance crisis of loop quantum gravity.
\end{abstract}

\section{Introduction}

Canonical gravity describes the 4-dimensional, generally covariant structure
of space-time by canonical fields defined on the slices of a spatial
foliation. Evolution of these fields in time as well as transformations
between different foliations are described by the geometrical structure of
hypersurface deformations. In a canonical theory, these transformations are
generated by certain phase-space functions, the diffeomorphism and Hamiltonian
constraints. In spherically symmetric models, which will be considered here,
the full set of constraints can be written as $D[M]$ and $H[N]$ with arbitrary
spatial functions $M$ (of density weight $-1$) and $N$. The constraint
equations $D[M]=0$ and $H[N]=0$, valid for any $M$ and $N$, restrict the
phase-space degrees of freedom, given by the spatial metric and its momentum
related to extrinsic curvature.

At the same time, the constraints generate (i) time evolution,
\begin{equation}
  {\cal L}_{t(N,M)}f=\{f,H[N]+D[M]\}
\end{equation}
for a phase-space function $f$ along a time-evolution vector field
$t^a=Nn^a+Ms^a$ in space-time with the unit normal $n^a$ to a spatial slice
and the tangent vector field $s^a=(\partial/\partial x)^a$ within the radial
manifold (with coordinate $x$) of a spatial slice, and (ii) gauge
transformations
\begin{equation}
  \delta_{\xi(\eta,\epsilon)}f=\{f,H[\eta]+D[\epsilon]\}
\end{equation}
along a space-time vector field
\begin{equation} \label{xi}
  \xi^a=\eta n^a+\epsilon s^a
\end{equation}
where $\epsilon$, like $M$, has density weight $-1$.

The reference to normal and tangential directions relative to a foliation
implies crucial differences between the mathematical formulation of
hypersurface deformations in canonical gravity and the more common formulation
of general covariance in terms of space-time tensors. In space-time, vector
components $\xi^a$ transform, by definition, in such a way that
$\xi^a\partial/\partial x^a$ determines a unique direction independent of
coordinate choices. Similarly, the spatial vector
$\epsilon s^a=\epsilon \partial/\partial x$ defines a coordinate-independent
direction because a scalar of density weight $-1$ in one dimension transforms
like a 1-form dual to $\partial/\partial x$. The normal deformation, however,
cannot be introduced in this way because the canonical setting does not
provide a time coordinate or the corresponding $\partial/\partial
t$. Moreover, even if such a coordinate could be introduced by hand, for
instance by using $t$ merely as a parameter as it also appears in Hamilton's
equations, it would be impossible to endow $\eta$ with a density weight $-1$
in the time direction because, canonically, there is no time manifold. The
only alternative is given by the procedure that has been used since
\cite{Katz,ADM} and formalized in \cite{Regained}: The normalization of $n^a$
as a unit vector (with respect to the space-time metric, which is
available in the canonical setting through the spatial metric on a slice as
well as lapse $N$ and shift $M$) associates a unique normal displacement to
any given function $\eta$ (without density weight).

The normal can be made unit only by reference to the metric, which provides
some of the canonical degrees of freedom. The geometrical meaning of normal
hypersurface deformations and their commutators depend on the spatial
metric, resulting in structure functions in the canonical bracket
relations. As a consequence, the canonical symmetries do not form a Lie
algebra. This property is responsible for several complications well-known in
attempts of canonical quantizations of the theory, starting with
\cite{NonHerm}. It also makes it harder to develop suitable mathematical
structures for transformations generated by the constraints, in particular in
an off-shell manner when one does not insist on solving the constraint
equations. In \cite{Regained}, for instance, it was shown that a direct
composition of transformations generated by the constraints is meaningful in
the sense of path independence (a notion introduced in there) only on-shell.

The full structure of transformations is nevertheless required for general
covariance to be implemented properly in the solutions of a canonical theory
of gravity, in particular one that has been quantized, modified or deformed by
new physical effects.  While the restricted on-shell behavior may be easier to
handle, the off-shell structure is important to make sure that the theory has
a well-defined space-time structure, independently of the dynamics. Only in
this case can the theory be considered a {\em geometrical} effective theory of
some deeper and as yet unknown quantum space-time, just as different dynamical
versions of gravity given by higher-curvature effective actions make use of
the same Riemannian form of space-time. Because of its importance for
covariance and the classification of meaningful effective theories, we will
review the structure of hypersurface deformations in the beginning of our
first section below, combining classic results from gravitational physics with
more recent mathematical developments \cite{ConsAlgebroid,ConsRinehart}.

We will focus on aspects of hypersurface deformations of importance for a
suggested simplification of the hypersurface-deformation brackets in
spherically symmetric models, given by a partial Abelianization
\cite{LoopSchwarz}, but our statements will apply also to a variety of other
reformulations that rely on phase-space dependent lapse and shift. Analyzing a
partial Abelianization in the context of hypersurface deformations, we will
show that this construction captures only a certain subset of these
transformations and, upon modification or quantization, does not guarantee
that invariance under hypersurface deformations or general covariance are
still realized. This conclusion may be surprising because, at first sight, a
partial Abelianization appears to implement the same number of symmetry
generators as standard hypersurface deformations and uses only a linear
redefinition of the generators. However, the coefficients of these linear
redefinitions are phase-space dependent, complicating their mathematical
description \cite{ConsAlgebroid,ConsRinehart}. (Heuristically, phase-space
dependent linear redefinitions of the generators introduce new structure
functions or modify existing ones.) It is then a non-trivial question whether
the redefinitions can be inverted. If they cannot be inverted, the redefined
theory is not invariant under full hypersurface deformations and its solutions
violate general covariance. An additional construction is therefore needed in
a partially Abelianized model (or other reformulations of standard
hypersurface deformations) in order to recover all space-time
transformations. As shown by explicit examples, this is not always possible if
the generators have been modified by quantum corrections.

A recent paper \cite{LoopSchwarzCov} claims that it may be possible to realize
general covariance in partial Abelianizations of spherically symmetric models
with different types of quantum modifications, such as a spatial
discretization. The claim is not accompanied by a successful reconstruction of
hypersurface deformations and instead relies on a technical and so far
incomplete case-by-case study of quantities that should be invariant in a
covariant theory. Using our results about general hypersurface deformation
structures, we will explain why the covariance claims of \cite{LoopSchwarzCov}
cannot hold.

\section{Hypersurface deformations}

Space-time vector fields with their standard Lie bracket generate the Lie
algebra of diffeomorphisms. Similarly, the transformations generated by the
canonical constraints form an algebraic structure. They are labeled by the
components $\eta$ and $\epsilon$ of a vector field $\xi$ used in (\ref{xi}) in
a basis $(n^a,s^a)$ adapted to a spatial foliation, rather than a coordinate
basis. Their commutators
\begin{eqnarray} \label{deltadelta}
  &&\delta_{\xi_2}(\delta_{\xi_1}f)-\delta_{\xi_1}(\delta_{\xi_2}f)\nonumber\\
  &=&
  \{\{f,H[\eta_1]+D[\epsilon_1]\}, H[\eta_2]+D[\epsilon_2]\}-
  \{\{f,H[\eta_2]+D[\epsilon_2]\},
                                                                     H[\eta_1]+D[\epsilon_1]\}\nonumber\\ 
  &=&
  \{f, \{H[\eta_1]+D[\epsilon_1], H[\eta_2]+D[\epsilon_2]\}\}
\end{eqnarray}
are determined by Poisson brackets
$\{H[\eta_1]+D[\epsilon_1], H[\eta_2]+D[\epsilon_2]\}$ of the constraints
(using the Jacobi identity). Because the unit normal $n^a$ is normalized by
using the space-time metric, including the spatial components $q_{ab}$ on a
slice, the brackets of two canonical gauge transformations
\cite{DiracHamGR,Katz,ADM} turn out to depend on the metric. In spherically
symmetric models, in which the radial part of the metric is determined by a
single function, $q$ (of density weight 2), we have
\begin{equation} \label{HDHD}
\{H[\eta_1]+D[\epsilon_1], H[\eta_2]+D[\epsilon_2]\}= H[\epsilon_1
\eta_2'- \epsilon_2\eta_1']+
D[\epsilon_1\epsilon_2'-\epsilon_2\epsilon_1'+
q^{-1}(\eta_1\eta_2' -\eta_2\eta_1')] \,.
\end{equation}
In general, the metric components are spatial functions independent of the
components $\eta$ and $\epsilon$ that label different gauge
transformations. Unlike the Lie bracket of two space-time vector fields, the
bracket of two pairs $\delta_{\xi_i}$, $i=1,2$, implied by the Poisson
bracket (\ref{HDHD}) does not form a Lie algebra because coefficients determined
by spatial fields $q_{ab}$ or $q$ cannot be considered structure constants.

\subsection{Algebroids}

Instead, the brackets have structure functions or, in a suitable mathematical
formulation, form the higher algebraic structure of an $L_{\infty}$-algebroid
rather than a Lie algebra \cite{StrongHomotopy,ConsHomology,ConsBFV}. An
$L_{\infty}$-algebroid is defined as a vector bundle over a base manifold $M$
with fiber $F$ and bracket relations on bundle sections together with suitable
anchor maps that map bundle sections to objects in the tangent bundle of $M$.
A Lie algebroid \cite{Pradines}, for instance, has a Lie bracket
$[\cdot,\cdot]$ on its sections and an anchor $\rho$ that maps (as a
homomorphism) bundle sections to vector fields on the base manifold, such that
the Lie bracket of vector fields is compatible with the algebroid bracket. The
anchor map also appears in the Leibniz rule
\begin{equation}
  [s_1, f s_2]= f[s_1,s_2]+ s_2 {\cal L}_{\rho(s_1)} f
\end{equation}
where $s_1$ and $s_2$ are sections and $f$ is a function on the base
manifold. The anchor brings abstract algebraic relations on bundle sections in
correspondence with geometrical transformations as vector fields on the base
manifold. While an anchor that maps any section to the zero vector field is
always consistent with the Lie-algebroid axioms (in which case the Lie
algebroid is a bundle of Lie algebras given by the fibers), non-trivial
transformations on the base require a larger image of the anchor. A Lie
algebroid with a non-trivial anchor generalizes bundles of Lie algebras.  Yet
more generally, and in particular in the case of structure functions, the
brackets of bundle sections obey the axioms of an $L_{\infty}$-algebra, a
generalized form of a Lie algebra in which the Jacobi identity is not required
to hold strictly.

The introduction of the base manifold makes it possible to formalize brackets
with structure functions in terms of an $L_{\infty}$-algebroid. In particular
for gravity, the base manifold is (a suitable extension \cite{ConsRinehart})
of the canonical phase space, given by the spatial metrics and momenta related
to extrinsic curvature. The fibers are parameterized by the components $\eta$
and $\epsilon$ of a gauge transformation. A section is then an assignment of
spatial functions $\eta$ and $\epsilon$ to any metric (or a pair of a metric
and its momentum). In this way, the $q$-dependent structure function in
(\ref{HDHD}) finds a natural home as a bracket of sections over the space of
metrics (and momenta).

Constant sections, given by pairs of $\eta$ and $\epsilon$ that are functions
on space but do not depend on the phase-space degrees of freedom, have a
bracket, implied by (\ref{deltadelta}), that can be realized as a special case
of sections of a Lie algebroid \cite{ConsAlgebroid}. General, non-constant
sections of this Lie algebroid have a bracket that may differ from what
hypersurface deformations would suggest.  Non-constant sections over phase
space, discussed in more detail in \cite{ConsRinehart}, either violate some of
the Lie-algebra relations on sections (in the controlled way of a specific
$L_{\infty}$-structure, as it follows from a BV-BFV extension of general
relativity \cite{Schiavina,BVBFVGR}) or require a base manifold that extends
the phase space of canonical gravity in a way that is not smooth. (The latter
can be formulated by using the notion of a Lie-Rinehart algebra
\cite{Rinehart} in which functions on the base manifold are replaced with a
suitable commutative algebra.

Phase-space dependent functions $\eta$ and $\epsilon$ are also important for
physics. They are often considered in specific gravitational
applications, as in the simple case of cosmological evolution written in
conformal time where the lapse function equals the scale factor, a metric
component. More importantly for our purposes, the partial Abelianization of
\cite{LoopSchwarz} relies on an application of phase-space dependent
$\epsilon$ and $\eta$. Hypersurface deformations with such non-constant
sections form a Lie algebroid only on-shell \cite{ConsRinehart} when the
constraints are solved. The partial Abelianization is therefore able to
describe the solution space to all constraints and its covariance
transformations, but it is not guaranteed that it correctly captures off-shell
transformations which are relevant for general covariance.

Since the standard derivation of the brackets (\ref{HDHD}) assumes that $\eta$
and $\epsilon$ are not phase-space dependent, the general brackets must be
extended by additional terms that, heuristically, result from Poisson brackets
of constraints with phase-space dependent $\eta$ and $\epsilon$. (A complete
derivation is based on the BV-BFV analysis of \cite{Schiavina,BVBFVGR}.) The
Poisson bracket of two diffeomorphism constraints, for instance, can still be
written in the compact form
\begin{equation} \label{DDfull}
  \{D[\epsilon_1],D[\epsilon_2]\}=
  D[\epsilon_2\epsilon_1'-\epsilon_1\epsilon_2']
\end{equation}
but with an application of the chain rule in the derivatives. Similarly, the
mixed Poisson bracket of a Hamiltonian and a diffeomorphism constraint in
general form reads
\begin{equation}\label{HDfull}
  \{H[\eta],D[\epsilon]\}= H[-\epsilon\eta']+ D[\eta{\cal L}_n\epsilon]
\end{equation}
where the normal derivative ${\cal L}_n$ of a spatial function is defined by
the Poisson bracket with the Hamiltonian constraint,
$\eta_1{\cal L}_n\eta_2= \{H[\eta_1],\eta_2\}$. For two Hamiltonian
constraints, we have the Poisson bracket
\begin{equation} \label{HHfull}
  \{H[\eta_1],H[\eta_2]\}=
  D[q^{-1}(\eta_1\eta_2'-\eta_2\eta_1')]+
  H[\eta_1{\cal L}_{n}\eta_2- \eta_2{\cal L}_n \eta_1]\,.
\end{equation}
In general, the extra terms implied by phase-space dependent $\eta$ and
$\epsilon$, such as those in
$\epsilon'=\partial_x\epsilon+(\partial_xq_i)(\partial_{q_i}\epsilon)+
(\partial_xk_i)(\partial_{k_i}\epsilon)$ summing over the two independent
components $q_i$, $i=1,2$, of a spherically symmetric spatial metric as well
as two components $k_i$ of extrinsic curvature, introduce further structure
functions, such as $\partial_xq_i$ and $\partial_xk_i$, that depend on the
metric as well as its momenta.

While these Poisson brackets illustrate the additional complications
encountered with phase-space dependent $\epsilon$ and $\eta$, they do not
immediately show the algebraic nature of general non-constant sections of
hypersurface deformations. In particular, Poisson brackets do not directly
mirror relevant $L_{\infty}$-structures. In our following discussion, we will
not need the full algebraic structure and instead perform a comparison of
different versions of constant and non-constant sections in gravitational
applications.

\subsection{Partial Abelianization}

As noticed in \cite{LoopSchwarz}, certain linear combinations of $H[\eta]$ and
$D[\epsilon]$ have vanishing Poisson brackets in spherically symmetric
models. In order to specify these combinations, we have to refer to explicit
variables that determine the spatial metric and its momenta. Following
\cite{SphSymm,SphSymmVol,SphSymmHam}, this is conveniently done in triad variables
$(E^x,E^{\varphi})$ such that the spatial metric is given by the line element
\begin{equation}
  {\rm d}s^2= \frac{(E^{\varphi})^2}{E^x} {\rm d}x^2+ E^x ({\rm d}\vartheta^2+
  \sin^2\vartheta{\rm d}\varphi^2)
\end{equation}
in standard spherical coordinates. (For our purposes, it is sufficient to
assume $E^x>0$, fixing the orientation of the triad.) The triad components are
canonically conjugate (up to constant factors) to components of extrinsic
curvature, $(K_x,K_{\varphi})$, such that
\begin{equation}
  \{K_x(x),E^x(y)\}= 2G\delta(x,y)\quad,\quad
  \{K_{\varphi}(x),E^{\varphi}(y)\}=G\delta(x,y)
\end{equation}
with Newton's constant $G$. (We keep a factor of two in the first relation. As
implicitly done in \cite{LoopSchwarz,LoopSchwarzCov}, this factor can easily
be eliminated by a rescaling of $K_x$. Since this procedure would not affect
the main equations and conclusions shown below, we do not make use of this
rescaling and instead keep the original components of extrinsic curvature.)

The delta functions disappear in Poisson brackets of integrated (smeared)
expressions, resulting in well-defined brackets. In particular, the
diffeomorphism constraint
\begin{equation}\label{D}
 D[M] = \frac{1}{G} \int{\rm d}x M(x) \left(-\frac{1}{2}(E^x)'K_x+K_{\varphi}'
   E^{\varphi}\right)\,,
\end{equation}
and Hamiltonian constraint
\begin{equation} \label{H}
 H[N]=\frac{-1}{2G}\int{\rm d}x N(x) \left(|E^x|^{-1/2}
   E^{\varphi}K_{\varphi}^2+
2 |E^x|^{1/2} K_{\varphi}K_x
+ |E^x|^{-1/2}(1-\Gamma_{\varphi}^2)E^{\varphi}+
2\Gamma_{\varphi}'|E^x|^{1/2}\right)
\end{equation}
where $\Gamma_{\varphi}=-(E^x)'/(2E^{\varphi})$ have Poisson brackets
\begin{eqnarray}
 \{D[M_1],D[M_2]\} &=& D[M_1M_2'] \label{DD}\\
 \{H[N],D[M]\} &=& -H[MN'] \label{HD}\\
 \{H[N_1],H[N_2]\} &=& D[E^x(E^{\varphi})^{-2}(N_1N_2'-N_2N_1')] \label{HH}
\end{eqnarray}
(for spatial functions $M_i$ and $N_i$, $i=1,2$, that do not depend on the
phase-space variables) of the correct form for hypersurface deformations in
spherically symmetric space-times.

Simple algebra and integration by parts shows that the linear combinations
\begin{equation} \label{CHD}
  C[L]=H[(E^x)'(E^{\varphi})^{-1} \smallint E^{\varphi}L{\rm
    d}x]-2D[K_{\varphi}\sqrt{E^x}(E^{\varphi})^{-1} \smallint E^{\varphi}L{\rm d}x]\,,
\end{equation}
where $\int E^{\varphi}L{\rm d}x$ is understood as a function of $x$ obtained
by integrating $E^{\varphi}L$ from a fixed starting point up to $x$, have zero
Poisson brackets with one another for different $L$:
\begin{equation} \label{CC}
  \{C[L_1],C[L_2]\}=0
\end{equation}
for all functions $L_1$ and $L_2$ on a spatial slice. To see this, it is
sufficient to notice that the combination eliminates any dependence on $K_x$
and on spatial derivatives of $E^{\varphi}$. The antisymmetric nature of the
Poisson bracket then implies that it must vanish. Explicitly, the new
combination of constraints takes the form
\begin{equation} \label{CL}
 C[L]= -\frac{1}{G}
\int{\rm d}x L(x) E^{\varphi}\left(\sqrt{|E^x|}
 \left(1+K_{\varphi}^2- \Gamma_{\varphi}^2\right)+{\rm const.}\right)\,.
\end{equation}
A free constant appears because a constant $\int E^{\varphi}L{\rm d}x$ implies
a non-vanishing lapse function in (\ref{CHD}), and therefore a non-trivial
constraint, but corresponds to a vanishing $E^{\varphi}L$ in (\ref{CL}). The
new constraint $C[L]$ therefore constrains one degree of freedom less than the
original $H[N]$. The free constant in (\ref{CL}) can be determined through
boundary conditions, which would also restrict the lapse functions allowed in
gauge transformations.

At first sight, it seems that the partial Abelianization eliminates structure
functions from the brackets and may simplify quantization and the preservation
of symmetries and therefore covariance. However, the importance of
metric-dependent structure functions in the standard brackets, which make sure
that deformations are defined with respect to a unit normal that is in fact
normalized, raises the question of whether an elimination of these structure
functions and their metric dependence by redefined generators can still
capture the full picture of general covariance. To answer this question, it is
instructive to place the partial Abelianization of the brackets in the context
of the hypersurface-deformation structure. Several features of the full
mathematical construction are then relevant.

First, the integration of $E^{\varphi}L$ required to define $C[L]$ as a
combination of $H[N]$ and $D[M]$ may seem unusual, but while this means that
the relevant $N$ and $M$ are non-local in space, they are local within both
the fiber (spatial functions $N$ and $M$) and the base (the gravitational
phase space with independent functions $E^x$, $E^{\varphi}$, $K_x$ and
$K_{\varphi}$ or a suitable extension) that may be used to construct a
corresponding $L_{\infty}$-algebroid. The combination (\ref{CHD}) therefore
defines an admissible set of sections.

Secondly, while the section defined by (\ref{CHD}) makes use of phase-space
dependent $N$ and $M$ in the Hamiltonian and diffeomorphism constraints, which
are therefore not constant over the base manifold, an Abelian bracket
(\ref{CC}) is obtained only for functions $L_1$ and $L_2$ that do not have the
full phase-space dependence allowed for general sections. In particular, if
$L_1$ or $L_2$ are allowed to depend on $(E^{\varphi})'$ or $K_x$, the bracket
$\{C[L_1],C[L_2]\}$ no longer vanishes, and it can then have structure
functions. Partial Abelianization is therefore obtained for a restricted class
of sections, defined such that $L$ does not depend on $(E^{\varphi})'$ and
$K_x$ (while it may still have an unrestricted spatial dependence). If $L$
does not depend on $(E^{\varphi})'$ and $K_x$ but on the other independent
phase-space variables, $K_{\varphi}$ as well as $E^x$ or on $E^{\varphi}$ but
not its derivatives, the bracket $\{C[L_1],C[L_2]\}$ remains zero, but there
are then structure functions in the bracket of $C[L]$ with the diffeomorphism
constraint, analogously to (\ref{HDfull}). Therefore, structure functions are
eliminated from the brackets only for a restricted class of sections. This
observation raises the question whether full covariance can still be realized.

A restriction to constant sections over the base manifold is not unusual, for
certain purposes. A similar assumption is made in the standard form
(\ref{DD})--(\ref{HH}) of hypersurface-deformation brackets, in which case the
original $N$ and $M$ are often assumed to be constant over the base (while
their spatial dependence remains unrestricted). There is, however, a crucial
difference between assuming constant $N$ and $M$ over the base and assuming
constant $L$ over the base: In the former case, allowing for non-constant
sections produces additional terms in the brackets, shown in (\ref{DDfull}),
(\ref{HHfull}) and (\ref{HDfull}), that follow directly from an application of
the product rule of Poisson brackets. The partial Abelianization, however,
relies on cancellations between different structure functions in the original
brackets that are no longer realized once non-constant sections with
phase-space dependent $L$ are allowed.

In particular, allowing for phase-space dependent $L$ and $M$ in the
$(D[M],C[L])$ system makes the transformation from $(N,M)$ to $(M,L)$
invertible. It is then possible to write the original $H[N]$ as a combination
of $D[M]$ and $C[L]$ in the partial Abelianization, regaining the full
non-Abelian brackets with metric-dependent structure functions. Restricting
the system to phase-space independent $L$, by contrast, implies that the
transformation from the original hypersurface-deformation structure to the
brackets of $D[M]$ and $C[L]$ is not invertible. It is then unclear whether
hypersurface deformations and general covariance can be recovered from a
partial Abelianization, in particular if the latter has been modified by
quantum corrections.

\subsection{Modified deformations}

It has been known for some time \cite{JR,LTBII,HigherSpatial} that spherically
symmetric hypersurface deformations can be modified consistently, maintaining
closed brackets while modifying the structure functions. The dependence on
$K_{\varphi}$ in (\ref{H}) can be generalized to
\begin{equation} \label{Hf}
 H[N]=\frac{-1}{2G}\int{\rm d}x N(x) \left(|E^x|^{-1/2}
   E^{\varphi}f_1(K_{\varphi})+
2 |E^x|^{1/2} f_2(K_{\varphi})K_x
+ |E^x|^{-1/2}(1-\Gamma_{\varphi}^2)E^{\varphi}+
2\Gamma_{\varphi}'|E^x|^{1/2}\right)
\end{equation}
where $f_1$ and $f_2$ are functions of $K_{\varphi}$ related by
\begin{equation} \label{f1f2}
  f_2(K_{\varphi})=\frac{1}{2} \frac{{\rm d}f_1(K_{\varphi})}{{\rm
      d}K_{\varphi}}\,.
\end{equation}
If this equation is satisfied, the bracket of two Hamiltonian constraints is still closed,
\begin{equation} \label{HHbeta}
  \{H[N_1],H[N_2]\} = D[\beta(K_{\varphi})E^x(E^{\varphi})^{-2}(N_1N_2'-N_2N_1')]
\end{equation}
for phase-space independent $N_1$ and $N_2$. In this bracket, $D[M]$ is the
unmodified diffeomorphism constraint, but the structure function is multiplied
by a new factor of
\begin{equation} \label{beta}
  \beta(K_{\varphi})= \frac{{\rm d}f_2(K_{\varphi})}{{\rm d}K_{\varphi}}=
  \frac{1}{2} \frac{{\rm d}^2f_1(K_{\varphi})}{{\rm d}K_{\varphi}^2}\,.
\end{equation}
Additional terms in the bracket for non-constant sections follow immediately
from the product rule for Poisson brackets.

Similarly, the Abelianized constraint $C[L]$ can be generalized in its dependence
on $K_{\varphi}$, using the same function $f_1$ as before:
\begin{equation} \label{CLf}
 C[L]= -\frac{1}{G}
\int{\rm d}x L(x) E^{\varphi}\left(\sqrt{|E^x|}
 \left(1+f_1(K_{\varphi})- \Gamma_{\varphi}^2\right)+{\rm const.}\right)\,.
\end{equation}
Its brackets remain Abelian for phase-space independent $L$.  There is no
obvious term in $C[L]$ where the second function $f_2$ might appear or the
important consistency condition (\ref{f1f2}). It therefore seems easier to
modify (or quantize) the constraint $C[L]$ compared with $H[N]$. However, for
full hypersurface deformations and covariance to be realized in the modified
setting, we still have to make sure that the transformation from $(N,M)$ to
$(L,M)$ can be inverted. As shown in \cite{SphSymmCov}, this is possible only
if we also modify the transformation (\ref{CHD}) to
\begin{equation} \label{CHDf}
  C[L]=H[(E^x)'(E^{\varphi})^{-1} \smallint E^{\varphi}L{\rm
    d}x]-2D[f_2(K_{\varphi})\sqrt{E^x}(E^{\varphi})^{-1} \smallint E^{\varphi}L{\rm d}x]
\end{equation}
where $f_2$ obeys the same consistency condition with $f_1$, (\ref{f1f2}), as
derived from the modified Hamiltonian constraint. The partial Abelianization
and the original form of hypersurface deformations therefore imply
equivalent results, provided one makes sure that the transformation of
sections can be inverted. Only then can access to full hypersurface
deformations and covariance be realized.

\section{Non-covariant modifications of Abelianized brackets}

A recent paper \cite{LoopSchwarzCov} by Gambini, Olmedo and Pullin (GOP)
argues that general covariance can be realized in modified versions of
spherically symmetric models, for which a partial Abelianization of the
brackets plays a crucial role: As the abstract claims, ``We show explicitly
that the resulting space-times, obtained from Dirac observables of the quantum
theory, are covariant in the usual sense of the way --- they preserve the quantum
line element --- for any gauge that is stationary (in the exterior, if there is a
horizon).  The construction depends crucially on the details of the
Abelianized quantization considered, the satisfaction of the quantum
constraints and the recovery of standard general relativity in the classical
limit and suggests that more informal polymerization constructions of possible
semi-classical approximations to the theory can indeed have covariance
problems.''

These claims raise several questions. For instance, how can the construction
depend ``crucially on the details of the Abelianized quantization considered''
if a partial Abelianization is either completely equivalent to the non-Abelian
orignal version of hypersurface deformations (if the transformation is made
sure to be invertible) or gives access to only a subset of hypersurface
deformations (if the transformation is not invertible owing to a restriction
to a subset of sections)?

A closer inspection of technical calculations performed by GOP shows that
spherically symmetric hypersurface deformations are, in fact, violated in the
construction. GOP use two different kinds of modifications, a generalized
dependence of $C[L]$ on $K_{\varphi}$ of the form (\ref{CLf}), and a spatial
discretization of phase-space functions and their derivatives. Because the
authors use a certain combination of solutions to the constraints and
gauge-fixing conditions, it turns out that only the latter modification
survives in the final expressions for line elements that are supposed to be
invariant.

However, also the former (a generalized dependence on $K_{\varphi}$) is relevant
because, as we have seen, the correct form of a modification must appear in
two different places, in the constraint $C[L]$ and in the transformation back
to unrestricted hypersurface deformations. These two appearances are clear but
somewhat implicit in \cite{LoopSchwarzCov}: The modified $C[L]$ is implied by
the modified solutions in equation (14) in \cite{LoopSchwarzCov} (or,
equivalently, (21) there, referring to the preprint version) where
$f_1(K_{\varphi})=\sin^2(\rho K_{\varphi})/\rho^2$ with a spatial function
$\rho$. The modified transformation back to unrestricted hypersurface
deformations is implied by equation (20) in \cite{LoopSchwarzCov} which in our
notation amounts to replacing $K_{\varphi}$ in (\ref{CHD}) with
$\sqrt{f_1(K_{\varphi})}$. Using the same function $f_1(K_{\varphi})$ is
crucial for the constructions in \cite{LoopSchwarzCov} because the partial
gauge fixing employed there replaces $\sqrt{f_1(K_{\varphi})}$ with a fixed
function on space (rather than phase space). The same gauge-fixing function is
then used in both places, in the constraint $C[L]$ or its solutions and in the
transformation back to unrestricted hypersurface deformations from which a
line element can be constructed. However, this construction, which is
equivalent to assuming $f_2(K_{\varphi})=\sqrt{f_1(K_{\varphi})}$ in
(\ref{CHDf}), violates the condition (\ref{f1f2}) required for unrestricted
hypersurface deformations to follow for the modified constraint. (For the
specific $f_1(K_{\varphi})$ considered by GOP, $f_2$ should have an additional
cosine factor, or equivalently have a doubled argument of the sine function.)
The constructions of \cite{LoopSchwarzCov} therefore violate hypersurface
deformations.

How can GOP then claim to have performed crucial steps toward demonstrating
general covariance in this setting? Unfortunately, much of the constructions
are obscured by an application of incompletely defined mixtures of gauge
fixings and idiosyncratic notions of observables. Here, it suffices to
highlight only a few of the shortcomings found in the GOP analysis. (For more
details, see \cite{LoopSchwarzCovComm}.)  Continuing with the replacement of
$\sqrt{f_1(K_{\varphi})}$ by a gauge-fixing function that depends only on
space, GOP replace any appearance of $\sqrt{f_1(K_{\varphi})}$ with
gauge-fixing functions (on space) derived from the classical solutions for
$K_{\varphi}$ in two specific slicings. Implicitly, the authors simply remove
the modification in this way because they indirectly equate
$\sqrt{f_1(K_{\varphi})}$ with $K_{\varphi}$, mediated by the gauge-fixing
function. As a result, they do not test how non-classical $f_1(K_{\varphi})$
can be consistent with covariance. It is also problematic that this step in a
rather careless gauge-fixing procedure replaces a phase-space function
$K_{\varphi}$ that does not Poisson commute with the constraints with a
spatial function that does obey this commutation property. The procedure
turns a $K_{\varphi}$-dependent expression for $E^{\varphi}$, obtained by
solving $C[L]=0$, into a function that Poisson commutes with $C[L]$. GOP then
call the result a Dirac observable, even though $E^{\varphi}$ is not gauge
invariant.

After replacing $K_{\varphi}$ with a spatial function, the resulting
expression for $E^{\varphi}$ still does not Poisson commute with the
diffeomorphism constraint and is therefore not a Dirac observable, even if
$K_{\varphi}$ could meaningfully be replaced.  The same expression for
$E^{\varphi}$ also depends on $E^x$, which is not a spatial invariant. Indeed,
unlike $C[L]$, the diffeomorphism constraint (\ref{D}) depends on $K_x$ and
therefore does not Poisson commute with $E^x$. GOP arrive at their conclusion
about $E^{\varphi}$ being a Dirac observable by misidentifying $E^x$ as a
Dirac observable because the (loop) quantization procedure they use
establishes a correspondence between an operator $\hat{E}^x$ and labels of a
spherically symmetric spin network state \cite{SymmRed,SphSymm} that are
unchanged by the spatial shifts of a finite diffeomorphism. However, having a
correspondence between a classical object, $E^x$, that is not a Dirac
observable and a quantum operator, $\hat{E}^x$, that is a Dirac observable
may indicate that the theory fails to have the correct classical limit. Since
this way of imposing the diffeomorphism constraint is directly inherited from
more general constructions in the full theory of loop quantum gravity
\cite{LoopRep,ALMMT}, the issues revealed by our analysis of
\cite{LoopSchwarzCov} might hint at deeper problems within the kinematics of
loop quantum gravity.

\section{Conclusions}

Our discussion of phase-space dependent coefficients in hypersurface
deformations has clarified a previously puzzling issue of partial
Abelianizations in spherically symmetric models: Is it possible for partial
Abelianizations to simplify the construction of quantum modifications of
hypersurface deformation generators and, at the same time, retain full access
to all transformations required for general covariance? We have shown that the
answer is negative. A simplified construction of modified generators is based
on the absence of structure functions in partially Abelianized brackets
obtained for a specific choice of phase-space dependent gauge generators
(lapse and shift functions). However, the partial Abelianization is maintained
only if the new generators are then restricted to be phase-space
independent. This condition renders the transformation from
hypersurface-deformation brackets to partially Abelian brackets
non-invertible. Access to unrestricted hypersurface deformations and general
covariance is therefore lost in a partially Abelianized setting. Consistent
modifications of the partially Abelian brackets then do not necessarily imply
consistent realizations of general covariance.

A recent paper \cite{LoopSchwarzCov} by Gambini, Olmedo and Pullin has
implicitly recognized this shortcoming and instead proposed to test general
covariance in a tedious case-by-case study of presumed invariants, beginning
with a discretized version of the line element. We have pointed out a specific
place (the choice of modification functions $f_1$ and $f_2$) where
hypersurface deformations are treated inconsistently in these constructions,
which may perhaps lead to improved versions of the transformations considered
by GOP. However, correcting this inconsistency requires an analysis of
unrestricted hypersurface deformations even in the partially Abelianized setting,
making sure that the transformation between these two versions of the brackets
can be inverted. It is therefore impossible to analyze covariance in isolation
from general hypersurface deformations, as proposed by GOP. No-go results
\cite{BlackHoleModels} for covariance in models of loop quantum gravity,
partially based on various analyses of modified hypersurface deformations,
therefore cannot be evaded by the constructions of GOP.

\section*{Acknowledgements}

The author thanks Michele Schiavina for discussions and Rodolfo Gambini,
Javier Olmedo and Jorge Pullin for sharing a draft of \cite{LoopSchwarzCov}.
This work was supported in part by NSF grant PHY-1912168.


\end{document}